\documentclass[epj]{svjour}
\usepackage{graphics}
\begin{document}
\title{Parity violation in nuclear systems}
\author{B. Desplanques
}                     
\institute{Laboratoire de Physique Subatomique et de Cosmologie 
(UMR CNRS/IN2P3--UJF--INPG),  \\
 F-38026 Grenoble Cedex, France}
\date{Received: date / Revised version: date}
%
\abstract{Parity violation in nuclear systems is reviewed. A few ingredients
relevant to the description of the parity-violating nucleon-nucleon force 
in terms of meson exchanges are reminded. Effects in nuclear systems are 
then considered. They involve $pp$ scattering, some complex nuclei 
and the deuteron system.
} 
\maketitle
%

\section{Introduction} \label{sec1}
A large number of parity-non-conserving (pnc) effects has been observed  
in various nuclear systems. While their expected size at low energy is 
of the order of $10^{-7}$ (for the amplitude), they can be strongly 
enhanced in some cases, due to the closeness of states with  opposite 
parities or the suppression of the regular transition. Thus, effects 
of the order of $10^{-1}$ have been measured in neutron-nucleus scattering 
in the vicinity of low energy p-wave resonances (see ref. 
\cite{Desplanques:1998} for some review). Qualitatively, such effects 
are understood. However, little quantitative information could be obtained 
on the pnc component of nucleon-nucleon ($NN$) forces expected to account 
for them. From the know\-ledge of this interaction, one can expect to learn 
about the pnc meson-nucleon coupling constants which they depend 
on and, thus, get information on the underlying hadronic physics. 
This one is complementary to the information that can be obtained 
from non-leptonic hyperon decays. It concerns in first place 
the $\pi NN$  coupling that has been at the center of many theoretical and 
experimental works. This one can be most easily compared to non-leptonic 
hyperon decay amplitudes. Another less fundamental but important motivation 
for the study of pnc nuclear effects is the necessity to determine 
the effective strength of the various pnc $NN$ amplitudes. These ones
can indirectly contribute to other pnc effects, especially in electron
scattering mainly discussed at this meeting. Though the effect is not large, 
its knowledge is required to determine the reliability of the information 
that is looked for in such high accuracy measurements. Some recent 
developments in the field are reviewed here.
 
The plan of the paper is as follows. In the second part, we briefly remind 
ingredients entering the  pnc $NN$ force, while emphasizing a few points of
interest for the following part devoted to pnc effects. The third section 
is concerned with pnc $pp$ scattering. This process is the only one that 
provides a calibration of the strength of pnc $NN$ forces at the present time. 
A few nuclear pnc effects in complex nuclei, especially in $^{18}$F and 
in $^{133}$Cs, are discussed in the fourth section. The fifth
section is devoted to pnc effects in the $np$ system, including the deuteron.
This particular field has been particularly active these last years. 
A conclusion and an outlook are presented in the sixth section. 

\section{PNC $NN$ potential: ingredients}  \label{sec2} \vspace{-2mm}
\begin{figure}[htb]
\begin{center} 
\resizebox{0.94\columnwidth}{!}{\includegraphics{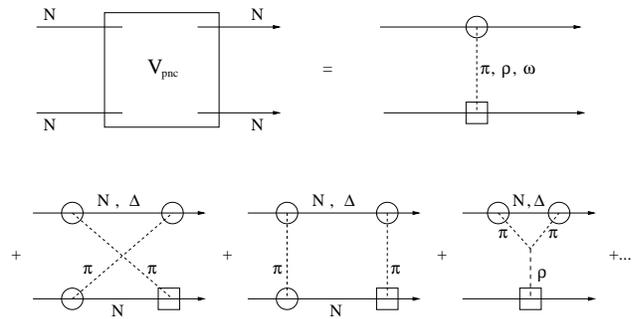}}
\end{center} 
\caption{Diagram representation of the pnc $NN$ interaction.
\label{fig1} }
\end{figure} 
The pnc $NN$ force is generally described as resul\-ting from meson exchanges,
$\pi$, $\rho$ and $\omega$. A diagrammatic re\-pre\-sentation is given in fig.
\ref{fig1}. One of the vertex, represented by a circle, corresponds to the
strong interaction. The other one, represented by a box, corresponds to the
weak, pnc, interaction.  As isospin is not conserved, there are many couplings
in some cases. They are: \\
- $h^1_{\pi}$, which governs the long range part of the force and necessarily
involves the $\Delta T =1$ component of the weak interaction \\
- $ h^{0,1,2}_{\rho}$:  $\Delta T =0,\;1,\;2$  \\
- $ h^{0,1}_{\omega}$:  $\Delta T =0,\;1$ \\
- $ h^{1'}_{\rho}$:  $\Delta T =1$, (different type $\rho NN$ 
coupling). \\
The vector-meson couplings determine the short-range part of the pnc $NN$ 
force. As is well known, the contribution of this part is sensitive to 
short-range correlations in the strong $NN$ interaction as well as to other
correlations. 

Many contributions going beyond the above ones have been considered in the
literature. They involve for instance two-pion exchanges displayed in fig.
\ref{fig1}, either with the same coupling as for the one-pion exchange 
or with the same coupling as for the $\rho $ exchange. In the last case, 
the $\rho $-exchange force acquires a longer range that could show up 
in the analysis of pnc effects in $pp$ scattering. Some discussion and 
references could be found in ref. \cite{Desplanques:1998}. 

At low energy, only gross features may be relevant. The $NN$ interaction 
can then be parametrized by five $S\leftrightarrow P$ $NN$ transition 
amplitudes \cite{Danilov:1972}:\\
- $ ^1S_0 \, \leftrightarrow\, ^3P_0$ , $\Delta T =0,1,2$, 
($pp$, $pn$ and $nn$ forces)\\
- $ ^3S_1 \,\leftrightarrow\, ^1P_1$ , $\Delta T =0$ ($pn$ force)\\
- $ ^3S_1 \,\leftrightarrow\, ^3P_1$ , $\Delta T =1$ ($pn$ force).\\
It was shown that this description could be extended to higher energy by
singularizing the pion-exchange contribution which, due to its long range, 
contributes sizeable  $P\leftrightarrow D$ transition amplitudes
\cite{Desplanques:1978}. Apart from the name, these works largely 
anticipated recent effective field-theory approaches \cite{Holstein:2004}, 
which also consider $P\leftrightarrow D$ transitions. 

Many works have been devoted to the pnc meson-nucleon couplings, which enter 
$NN$ interaction models. A large part of them, prior to the DDH work 
\cite{Desplanques:1980} or later, fit in this framework. Due to the lack 
of space, we again refer to ref. \cite{Desplanques:1998} for references 
and detailed discussion. We only present here some estimates and make 
a few pertinent remarks. The sample of results given in table \ref{table1}
corresponds to the predictions of two significantly different models for 
the most relevant couplings, $h^1_{\pi}$, $h^0_{\rho}$ and $h^0_{\omega}$. 
They are based on a quark model (DDH), partly updated, and a chiral soliton 
model  by Kaiser and Meissner (KM) \cite{Kaiser:1988} (see also ref. 
\cite{Meissner:1999}). Despite appearances, results turn out to be 
qualitatively similar. Discrepancies can be ascribed to the weight 
attributed to individual contributions in DDH. It is noticed that the 
dominant contribution to $h^1_{\pi}$ is produced by strange quarks, of
particular interest at this meeting while the consistency of this estimate
with the QCD sum rules ones remains an open problem. It was proposed to use 
a chiral quark model to make a new estimate (Lee {\it et al.} \cite{Lee:2004}, 
this conference). It is also noticed that DDH estimates, relying for a part 
on experimental data, should be less sensitive to ``rescattering effects'' 
evoked in the literature \cite{Zhu:2001} whereas  $h^0_{\omega}$ is likely 
to be negative.
\begin{table}[htb]
\caption{Meson-nucleon pnc coupling constants: a few estimates from different
works. The question mark at the last line indicates that the original value
could be actually close to 0.  \label{table1} }
\begin{tabular}{lccc}
  \hline \vspace{-3mm} \\
                        &  DDH (range)  &   DDH(``best'' )  &  KM  
  \\ [0.5ex] \hline  \vspace{-3mm}\\
 $10^7\;h^1_{\pi} \hspace*{0.3cm}$     & 0 $\leftrightarrow $ 11      
 &   4.6         & 0.2   \\ [1.ex] 
 $10^7\;h^1_{\pi} \hspace*{0.3cm}$     & 0 $\leftrightarrow $ 2.5     &  & 
 0.8 - 1.3   \\ [0.ex] 
 $  $     &  (update, $K=3$)    &  & 
  (with $\bar{s}s$)  \\ [1.ex] 
 $10^7\;h^0_{\rho} \hspace*{0.3cm}$     & -31 $\leftrightarrow $ 6    
  & -11           & -4   \\ [1.ex]
 $10^7\;h^0_{\omega} \hspace*{0.3cm}$   &  -10 $\leftrightarrow $ 6?     
 & -2           & -6  \\  [0.ex]    \hline
\end{tabular}
\end{table}
%

\section{Longitudinal asymmetry in $pp$ scattering} \label{sec3}
The low-energy longitudinal asymmetry in $pp$ scatering is the most important
benchmark in the field at present. A complete theoretical analysis can 
be done. It shows that measurements at 13.6 and 45 MeV are in complete 
agreement with each other, thus fixing the strength of the 
$^1S_0 \leftrightarrow\, ^3P_0$ pnc $pp$ transition amplitude. For a given 
description of the strong interaction model, the strength of a combination 
of   $h^{pp}_{\rho}$ and $h^{pp}_{\omega}$  couplings  or, in first 
appro\-ximation, the $h^0_{\rho}$ and $h^0_{\omega}$ couplings, can be 
obtained.  At higher energy, around 221 MeV, it was noticed that the 
contribution of the $^1S_0 \leftrightarrow\, ^3P_0$ transition amplitude 
was vanishing, providing a window to determine the  
$^3P_2 \leftrightarrow\, ^1D_2$ transition amplitude.  By combining 
this measurement with the low-energy one, contributions due 
to $\rho$ and $\omega$ exchanges  can thus be disentangled. 

\begin{figure}[htb]
\begin{center} 
\resizebox{0.94\columnwidth}{!}{\includegraphics{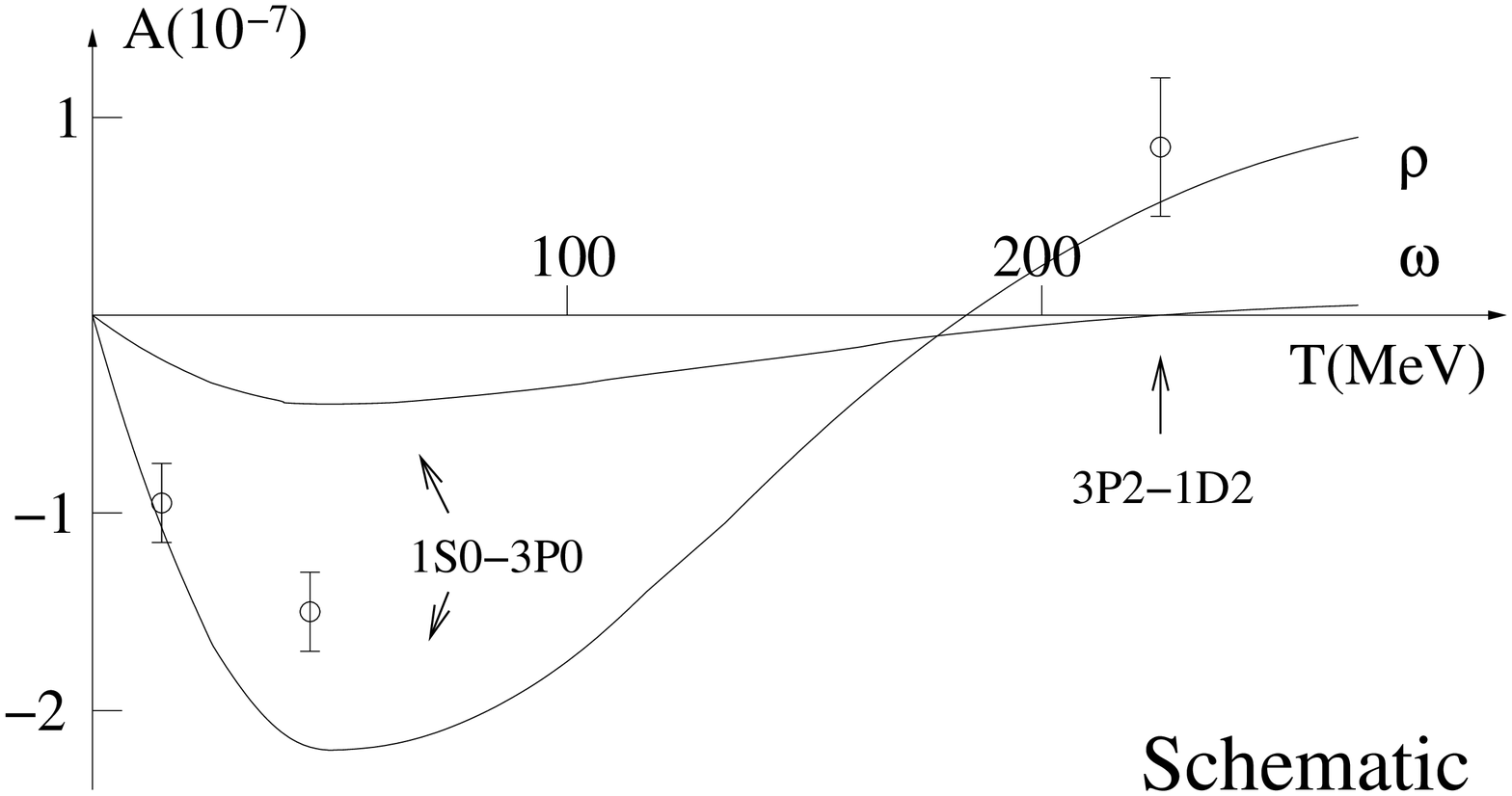}}
\end{center}
\caption{Schematic representation of $\rho$- and $\omega$-exchange 
contributions to the longitudinal asymmetry in $pp$ scattering. 
\label{fig3} }
\end{figure}  
A schematic representation of the two contributions with couplings close 
to the DDH ``best guess" ones, is given in fig. \ref{fig3}. It is seen that 
the $\omega$-exchange one has a ne\-gli\-gible contribution around 221 MeV  
while a $\rho$-exchange contribution alone is not doing badly. A better 
agreement is obtained by increasing the strength of this one and compensating 
for the overestimate at low energy by adding a $\omega$-exchange 
contribution with a sign opposite to the DDH ``best guess" or KM one.
This provides a simple explanation for the couplings obtained by Carlson 
{\it et al.} \cite{Carlson:2002}:
\begin{eqnarray}
&& h^{pp}_{\rho}= -22.3 \; 10^{-7},\;\;h^{pp}_{\omega}= +5.2 \; 10^{-7}\;\;
({\rm fit})
\nonumber \\
&&h^{pp}_{\rho}= -15.5 \; 10^{-7},\;\;h^{pp}_{\omega}= -3.0 \; 10^{-7},\;\;
({\rm DDH}).
\end{eqnarray}
The fit evidences a striking feature as the value for the  $\omega$ coupling 
has a sign  opposite to expectations and, thus, can point to missing 
ingredients in predictions. We however notice that the significance 
of the result is not strong (the $\rho$ alone is already giving a good account). 
A refined theoretical analysis and a more accurate measurement could be 
quite useful. It would be interesting to investigate for instance the role 
of a longer range $\rho$-exchange contribution mentioned in the second 
section.

\section{Parity-non-conservation in complex nuclei} \label{sec4}
Many pnc effects in complex nuclei have been measured and analyzed. 
Considered individually, it is however difficult to draw some conclusion 
from them. Moreover, when they are considered together, it is not rare 
that the information obtained from one process is at the limit to contradict 
that one from another process. Two effects ne\-ver\-theless deserve some 
attention: the circular polarization of photons emitted in the transition 
$0^{-} (1.08 \,{\rm MeV}) \rightarrow 1^{+}$(g.s.) in $^{18} $F and the $^{133}$Cs 
anapole moment. They are successively discussed in the following (references 
for both theory and experiment may be found in ref. \cite{Desplanques:1998}). 

The interest of the pnc effect in $^{18} $F is that the calculation of the 
relevant pnc nuclear matrix element ($0^{-} \leftrightarrow 0^{+}$) can 
be checked by looking at the first forbiden $\beta$ decay of the neighboring
nucleus, $^{18} $Ne. It implies the $\Delta T=1$ part of the weak interaction
and the measurement can thus provide information on the pnc $\pi NN $ 
coupling, $h^1_{\pi}$. From the experimental limit of $P_{\gamma}$, 
one gets the following upper limit:
\begin{equation}
 |h^1_{\pi}| \leq 1.3\, 10^{-7}.\label{hpi}
\end{equation}
This result is supported by the absence of effect  in two other processes in 
$^{21} $Ne and in $^{93}$Tc. In these cases, the contribution of the pion
exchange is {\it a priori} large. To agree with the upper experimental limit, 
one has first to assume that the coupling $h^1_{\pi}$ is not too large and, 
moreover, that the corresponding contribution be cancelled for a part 
by some isoscalar contribution (for $^{21} $Ne). Contrary to $^{18} $F, 
there is no available check on the relevant pnc nuclear matrix element. 
Accepting that this one be uncertain by up to a factor 3 would however 
give a limit on $h^1_{\pi}$ si\-mi\-lar to eq. (\ref{hpi}). 

The $^{133}$Cs anapole moment has been analyzed by different authors. 
To a large extent, this quantity involves a combination of the pnc $NN$ force 
close to that one go\-ver\-ning pnc effects in several odd-proton systems as
different as $p\, \alpha$ scattering, $^{19} $F, $^{41} $K,  $^{175} $Lu, 
$^{181} $Ta. At first sight, it appears that the above combination should be
two times larger for the anapole moment than for the other processes. The
discrepancy has the order of a typical uncertainty in the field but there was
some belief that the estimate in the first case could be less uncertain than
for other effects (for a part, it involves a long-range operator). 
On the other hand, the effects in the other odd-proton systems overdetermine 
the above combination of parameters. If these last processes are ignored, 
it appears that a large value of $h^1_{\pi}$, of the order of $1\, 10^{-6}$, 
at the upper limit of the original DDH range, is needed. This is inconsistent 
both with the upper limit, eq. (\ref{hpi}), and the DDH updated range. 
We notice that the last calculation of the anapole moment \cite{Haxton:2001} 
relies on an approximation that allows for an improved calculation in one 
respect but implies some contribution from orbitals below the Fermi level 
with a wrong sign in another respect. A correct account of these ones could 
enhance the theoretical estimate but will not reach a factor 2. The validity 
of a similar approximation, which omits 3-body terms, was discussed 
in ref. \cite{Desplanques:1973}.
 
\section{Parity-non-conservation in the deuteron} \label{sec5}
Most recent pnc studies in nuclear systems have concentrated on 
the $np$ system (deuteron and scattering state). This emphasis is largely 
motivated by both the feasibi\-li\-ty of the corresponding experiments 
in a near future (see Stiliaris's talk at this conference) and a safer 
interpretabi\-li\-ty of possible effects. These ones include the
photon-emission  asymmetry in the thermal-energy radiative capture of polarized 
neutrons by protons, $\vec{n} +p \rightarrow d+\gamma$
\cite{Kaplan:1999,Desplanques:2001,Hyun:2001,Schiavilla:2003}, presently
performed at LANSCE, 
the asymmetry in the deuteron photo-disintegration depending on the photon 
helicity \cite{Khriplovich:2001,Liu:2004,Fujiwara:2004,Schiavilla:2004}, 
which could be performed at JLab, IASA, SPring-8, $\cdots$, 
the deuteron anapole moment 
\cite{Savage:1998,Khriplovich:2000,Hyun:2003,Liu:2003}, 
the pnc deuteron electro-disintegration in relation with the SAMPLE experiment 
\cite{Liu:2003b,Schiavilla:2003}, and
the longitudinal asymmetry and the neutron-spin rotation in $np$ scattering 
\cite{Schiavilla:2004}.
Some earlier works could be quoted. The recent ones involve new methods 
(effective-field theories \cite{Kaplan:1999,Hyun:2001}), 
improved $NN$ interaction models (AV18+ $\cdots$ 
\cite{Hyun:2001,Liu:2003,Schiavilla:2003,Liu:2004,Schiavilla:2004}), 
more complete calculations (two-body currents 
\cite{Liu:2003b,Schiavilla:2003}), 
and increased attention to gauge invariance 
\cite{Hyun:2003,Liu:2003,Schiavilla:2004}. A few remarks are made below 
about these different works.

The earlier pion-exchange contribution to the asymmetry in 
the thermal-neutron radiative capture on protons, 
$\vec{n} +p \rightarrow d+\gamma$, 
\begin{equation}
A_{\gamma} = -0.11 \; h^1_{\pi},
\label{Agamma}
\end{equation}
is confirmed by recent estimates, indicating that the correction for a wrong
$^1S_0$ $np$ scattering length, was appropriately made. It is also found that the above
estimate results from a strong cancellation when a calculation is performed
without relying on the Siegert theorem \cite{Hyun:2001,Schiavilla:2003}. 
Amazingly, this weak interaction 
problem provides information on the accuracy of effective-field theory methods 
employed for the strong interaction. The  approach used in ref.
\cite{Kaplan:1999}, for instance, overestimes eq. (\ref{Agamma}) by 60\% 
at leading order (almost a factor 2 for comparable ingredients). 

On the basis of an estimate by Oka \cite{Oka:1983}, it was thought 
that the study of the photon-helicity dependence of the deuteron 
photo-disintegration cross section could provide an alternative 
way to determine the coupling $h^1_{\pi}$. This
motivated several works that disproved the above estimate  and its main 
conclusion \cite{Khriplovich:2001,Liu:2004,Fujiwara:2004,Schiavilla:2004}. 
An account  of the new results can be found in the Hyun's talk 
at this conference. For the inverse process near threshold 
(``Lobashov expe\-ri\-ment"), it should be
noticed that a circular polarization of photons as large as $1\; 10^{-7}$ 
is not excluded for some reasonable models of both the strong and the
weak $NN$ interaction \cite{Schiavilla:2004}. 

The deuteron anapole moment is largely academic as there is not much hope it
could be measured in a near future. It however provides a nice laboratory for
studying the implications of gauge invariance, which is essential for getting 
a consistent estimate of this quantity. A contribution required by chiral
symmetry \cite{Savage:1998}, absent in ref. \cite{Khriplovich:2000}, 
has thus been recovered in potential based approaches \cite{Hyun:2003}. 
On the other hand, this last work confirms the
conclusion obtained from the study of $A_{\gamma}$ about the reliability of 
some effective-field theories. It is likely that an alternative approach 
\cite{Savage:2001}, which is nothing but the one initiated by Danilov's work
\cite{Danilov:1972}, extended later on to higher energy 
\cite{Desplanques:1978}, should do better. 

In pnc-electron experiments performed on the deuteron, aiming at determining 
the contribution of strange quarks to nucleon form factors, there was 
some concern about the role of a nuclear pnc effect. This one was studied 
in two different works \cite{Liu:2003b,Schiavilla:2003}  
which showed that the effect, a few percents, would be negligible. Actually, 
the main role of pnc nuclear effects in this process (together with that one
involving the proton) is an indirect one. They allow one to put limits on
coupling constants that enter radiative corrections \cite{Zhu:2000}.

Parity-non-conservation in $np$ scattering has been recently revisited 
\cite{Schiavilla:2004}. The main feature evidenced by the new results is the
dominance of the pion-exchange contribution, as far as the DDH ``best guess" 
is used for the corresponding pnc coupling.

\section{Discussion and conclusion} \label{sec6}
Many low-energy pnc nuclear effects, involving mainly protons, are 
within expectations. However, one has often to be satisfied with 
discrepancies up to a factor 2. This is not enough to constrain 
the different pnc meson-nucleon couplings if one refers to a potential 
approach or the low-energy $NN$ scattering amplitudes if one rather relies on
the less ambitious approach represented by effective field theories. 

Most probably, the pnc $\pi NN$ coupling, $h^1_{\pi}$, is small and within 
the DDH updated range. Some processes could require a significantly larger 
value but, in our opinion, they have not the weight of the other ones 
that point to a small value. Concerning the vector meson-nucleon couplings,
there is a slight hint that the isoscalar $\omega$ one, $h^0_{\omega}$, could
have a sign opposite to expectations. This should motivate further studies to
confirm the hint on the one hand, to see whether this opposite sign is
conceivable.

An analysis in terms of couplings has some interest but, apart from 
the fact it assumes that multi-meson exchanges can be ignored, it does 
not necessarily provide a pertinent clue at which part of the pnc interaction 
is rather unconstrained. Looking at the various $NN$ scat\-te\-ring amplitudes can
thus represent a complementary view.  Among the five amplitudes required for 
the description of pnc effects at low energy, only one ($pp$) is determined 
with a good accuracy. From the study of odd-proton systems, and after 
removing the contribution of the $pp$ amplitude, a $pn$ amplitude involving 
``unpolarized" neutrons can be obtained. Being derived indirectly, 
from complex systems moreover, the accuracy of this amplitude is not 
as good as for the $pp$ one. For the three other amplitudes, which involve 
``polarized" neutrons (with ``unpolarized" protons, with ``unpolarized"
neutrons and with ``polarized" protons), only upper limits are known. 

To determine this sector of the pnc $NN$ interaction, appropriate experiments 
are heavily needed, preferentially with light systems where theoretical 
uncertainties are reduced.  The $np$ amplitude with ``polarized" neutrons 
is better studied in the neutron-spin rotation. The $nn$ amplitude 
could be obtained from the neutron-spin rotation in neutron-$\alpha$ 
scattering, after removing the previous contribution of the $np$ amplitude. 
The best process for determining the last $np$ amplitude, which involves 
both ``polarized" neutrons and ``polarized" protons, is the circular 
polarization of photons in the thermal neutron-proton
radiative capture (``Lobashov experiment"). Evidently, the asymmetry 
$A_{\gamma}$, already  mentioned, is part of the needed experiments. 
While it involves the difference in the two $np$ amplitudes with 
a ``polarized" neutron and an ``unpolarized" proton on the one hand, 
the inverse configuration on the other hand, it also allows one to get
information on the most debated pnc coupling,  $h^1_{\pi}$. 

\section*{Acknowledgments}
We are very grateful to C. H. Hyun and C.-P. Liu for a stimulating 
and fruitful collaboration.


\end{document}